\documentstyle[aps,twocolumn,prb,psfig]{revtex}
\newcommand{\bge}{\begin{equation}}
\newcommand{\ege}{\end{equation}}
\newcommand{\bga}{\begin{eqnarray}}
\newcommand{\ega}{\end{eqnarray}}

\input epsf
\begin{document}
\draft

\title{Higher anisotropic $d$-wave symmetry in cuprate superconductors}

\twocolumn[ 
\hsize\textwidth\columnwidth\hsize\csname@twocolumnfalse\endcsname 

\author{Haranath Ghosh}
\address{Department of Physics, University of Arizona, Tucson, AZ 85721, USA.}
\today
\maketitle
\begin{abstract}
We derive a pair potential from tight binding further neighbours attraction
that leads to superconducting gap symmetry similar to that
of the phenomenological spin
fluctuation theory of high temperature superconductors (Monthoux, Balatsky,
Pines, Phys. Rev. Lett. {\bf 67}, 3448).
We show that higher anisotropic $d$-wave than the simpliest
$d$-wave symmetry is one of the important ingredients responsible
for higher BCS characteristic ratio.
\end{abstract}
%
%
%
%
%
]
Pairing symmetry of the superconducting energy gap in high temperature
superconductors still remains an open problem after a decade of its discovery.
Various experimental results which lead to conflicting conclusions resulted
no concrete concencus to the theory of pairing mechanism for high $T_c$
superconductors. However, there are strong evidences that the pairing state of
the cuprate superconductors could be $d$-wave like; experimental
observations that are sensitive to the
phase \cite{1} and nodes \cite{2} of the gap, reported a sign
reversal of the order parameter supporting $d$-wave pairing.
On the other hand, a group of experiments on the same $YBa_2Cu_3O_7$ (YBCO)
material
indicate existence of a significant $s$-component \cite{3}; had YBCO been
a pure $d$-wave superconductor, it would be orthogonal to $s$-wave 
state of $Pb$
resulting zero Josephson supercurrent (while in experiment a well defined
$c$-axis current is seen) and
there are strong evidences that the electron doped $\rm Nd_{2-x}Ce_xCuO_4$
superconductors are s-wave type \cite{4}.
 There are also indications, both from theories and experiments, that
the high $T_c$ materials may have a mixed pairing symmetry ($e.g,$ $d\pm s$ or
$d +is/d_{xy}$ etc.) in presence of external magnetic field, magnetic impurity 
 \cite{5},
interface effects
etc. \cite{6}  In addition, there exists important clues that indicate
pairing state even in the bulk of the cuprates and in absence of magnetic field
may also have a mixed pairing state, with a minor component coexisting with
predominant $d$-wave ({\it i.e,} $d + e^{i \theta}\alpha$ scenario,
$\alpha = s, d_{xy}$) \cite{7}. The angle resolved photoemission spectroscopy
(ARPES) study by Kelley {\it et al}, provides strong indication that
$\rm Bi_2Sr_2CaCu_2O_{8+\delta}$ compound is $d$-wave like in the under- and
optimally doped regime whereas {\em not} a $d$-wave like in the slightly
overdoped but high-$T_c$ sample. Raman measurements
confirmed the unexpected behavior of gap symmetry (from predominant
$d_{x^2-y^2}$ in under- or optimally doped to anisotropic $s$-wave type
 in overdoped) by overdoping
$\rm Bi_2Sr_2CaCu_2O_{8+\delta}$ and almost similar phenomena is also found
in other high-$T_c$ material $\rm Tl_2Ba_2CuO_{6+\delta}$ \cite{8}.
Now, observation of any $s$ component will have stringent contraints 
on various
potential theories of high $T_c$ superconductors, such as, antiferromagnetic
spin fluctuation theory \cite{9}.
Because, antiferromagnetic spin fluctuation theory
leads to attraction in the $d$-wave channel and pair breaking in the $s$-wave
channel (but in a model calculation via spin fluctuation in heavy fermion
systems by Miyake {\it et al.} \cite{10} indicates possibility of anisotropic
$s$-wave pairing as well). On the other hand, pairing mechanisms based on
electron-phonon interactions, polarons etc. would be compatible with
pure $d$-wave, pure $s$-wave or an admixture of the two \cite{11}.

Therefore, it is evident from the above discussion that the symmetry of the
order parameter and the associated mechanism of pairing in high $T_c$ cuprates
are not at all clear, but is essential to have a first step
development towards an understanding to the mystry of pairing mechanism.
The spin fluctuation theory is one of the potential theories of high $T_c$
superconductivity which can account a number of anamolous properties
observed in cuprates. It is a phenomenological theory with a few small
parameters, like the
phenomenological form of the spin susceptibility $\chi (Q)$,
the magnetic coherence length $\xi$, the magnon frequency $\omega_{SF}$
extracted from Nuclear Magnetic Resonance (NMR) experiment \cite{nmr}.
The approximate momentum distribution of the superconducting energy gap
function obtained by the authors of ref. \cite{9} is,
\begin{equation}
\Delta_{SF} (k) = \Delta (0) (\cos k_x a - \cos k_y a)
\sum_N (\cos k_x a + \cos k_y a)^N
\end{equation}
This is not a {\em lowest order} $d$-wave symmetry as is usually considered in
the literature. It was Lenck and Carbotte, who first pointed out this fact
\cite{lenck}. They obtained the superconducting gap function by using BCS
theory with the phenomenological spin susceptibility as the pairing interaction
by using fast-Fourier-transform technique, without any prior assumption about
the symmetry of the gap function. They concluded that the gap structure
although have nodal lines along $k_x = k_y$ cannot have the simple
form of $\cos k_x a - \cos k_y a$ with $a$ the lattice parameter.
In a weak coupling theory language, in order to get a gap symmetry as
$\Delta_{SF} (k)$ one needs a pair potential which also has the same symmetry.
We show in a tight binding picture by considering higher neighbours attraction
that such potential is derivable up to third order term in equation (1). After
obtaining the pair potential we calculate the explicit structure of gap
function thus obtained and found to be similar to that of Lenck and Carbotte.
We also point out that such higher anisotropic $d$-wave
symmetry is key to understand larger $2 \Delta (k)_{max}/k_BT_c$ BCS ratio.

 In the spirit of the tight binding description assuming that the overlap
of orbitals in different unit cells is small, compared to the diagonal overlap
values, the matrix element $V(k,k^\prime)$ may be written as,
\begin{equation}
V(\vec q) = \sum_{\vec \delta} V_{\vec \delta} e^{i \vec q \vec R_{\delta}}
=V_{o}^r + 2\sum_{n=1,}^3 V_n (\cos q_x na + \cos q_y na)
\end{equation}
where $\vec R_{\delta} = \pm n a$ locates nearest neighbours and further
neighbours ; since we shall only be interested in the $d$-wave channel ({\it
i.e}, in a square lattice we are not considering 2nd, 5th etc.
neighbour matrix elements as it gives rise to $d_{xy}, s_{xy}$ channels).
Thus we get,
from the requirement of singlet pairing symmetry {\it i.e}, $\Delta (k)=
\Delta (-k)$,
\begin{equation}
V(k,k^\prime)=V_{o}^r + \sum_{n} V_n f_{k}^n f_{k^\prime}^n + \sum_{n} V_n
g_{k}^n g_{k^\prime}^n
\end{equation}
where $f_{k}^n (g_{k}^n) = \cos k_x na \mp \cos k_y na$, $V_{o}^r$ is the
on-site term (the label $r$ stands for repulsion, but could be attractive
as well giving rise to isotropic $s$ wave pairing) and the $3^{rd}$ term in
equation (3) responsible for extended s-wave pairing will be omitted from
further discussion. 

In deriving equations (2,3) we have taken into account attractions only along
the $x$ and $y$ axis neighbours. However, such attractive interaction
 between the $4^{th}$
neighbours also gives rise to unconventional $d$-wave pairing channel, the 
pair potential for the $4^{th}$ neighbour interaction that leads to
singlet $d$-wave pairing may be obtained as,
\begin{eqnarray}
V(k,k^\prime)& =& 2V_4(\cos k_x - \cos k_y)(1+2 \cos k_x \cos k_y)\times
\nonumber \\ &&
(k\to k^\prime) + 
 2 V_4 ( \cos k_x - \cos k_y)(2 \sin k_x \sin k_y)
\nonumber \\ && \times (k\to k^\prime)
\end{eqnarray}
where $V_4$ indicates strength of $4^{th}$ neighbour attraction.

Thus, taking into consideration just the $d$-wave channel, one obtains the
anisotropic pair potential as,
\begin{eqnarray}
V(k,k^\prime)&=&V_1f_kf_{k^\prime}+4V_3 f_kf_{k^\prime}g_{k}g_{k^\prime}
+2V_4f_k f_{k^\prime}[(1+d_{k_{xy}}) \nonumber \\ &&
\times(1+d_{k_{xy}^\prime})+
s_{k_{xy}}s_{k_{xy}^\prime}]  
+16V_6
f_kf_{k^\prime}\times  \nonumber \\ && 
(g_{k}^2-\frac{s_{k_{xy}}}{2}-\frac{3}{4}) 
(g_{k^\prime}^2-\frac{s_{k_{xy}^\prime}}{2}-\frac{3}{4}) 
 \nonumber \\ && \approx 
Vf_kf_{k^\prime}(1 +g_{k}g_{k^\prime}+g_{k}^2 g_{k^\prime}^2)
\end{eqnarray}
where we assumed $2V_1=4V_3=4V_4=16V_6$ =V (say) and considered only the 
appropriate contributing terms in $V_4$ and $V_6$ that leads to the
gap structure in equation (1) [in deducing the
$2^{nd}$ result of Eq (5)]. In equation (5) $V_1$, $V_3$, $V_4$, $V_6$
  represents
strength of attraction between the $1^{st}$, $3^{rd}$, $4^{th}$, $6^{th}$
  neighbours 
respectively and the momentum form factors are 
$d_{k_{xy}} = 2 \sin k_x \sin k_y$, $s_{k_{xy}} = 2 \cos k_x \cos k_y$,
$f_k = \cos k_x - \sin k_x$, $g_k = \cos k_x + \cos k_x$, with
$f_{k}^2 = (f_k)^2~ \& ~g_{k}^2 = (g_k)^2$.  The actual approximation involved
to get  the second result of the above equation are, 
\begin{eqnarray}
&&2V_4f_k f_{k^\prime}[(1+d_{k_{xy}})(1+d_{k_{xy}^\prime})+
s_{k_{xy}}s_{k_{xy}^\prime}]\sim \frac{V}{2}f_k f_{k^\prime} \nonumber \\
&& 
V_6 f_k f_{k^\prime}(4 g_{k}^2 - 2 s_{k_{xy}} -3)
(4 g_{k^\prime}^2 - 2 s_{k_{xy}^\prime}  -3) 
 \sim \nonumber \\
&&16 V_6 f_kf_{k^\prime}g_{k}^2 g_{k^\prime}^2
\end{eqnarray}
These approximations are used just to retain the form of the gap structure
(1), however, full form of the potential, the first result of equation (5)
will also be explicitly used. It turns out that the full potential leads to
results very close to that obtained in the spin fluctuation theory, whereas
the form given as
 equation (1) is just as an artefact of the approximations used
in the ref. \cite{9}.

 We shall show now that the pair-potential in the second result of 
equation (5) can produce the gap symmetry of the spin fluctuation theory
given in equation (1) up to
$3^{rd}$ order term {\it i.e}, $\Delta(k) = \Delta(0)f_k(1+g_{k}+g_{k}^2)$.
(Note, in a weak coupling BCS theory one would tend to think of a potential,
$V (k,k^\prime) = V F_k F_{k^\prime}$, where $ F_k = f_k(1+g_{k}+g_{k}^2)$
should be essential to produce the above gap symmetry).
Supposing that the pair-potential in the second result of (5) 
does produce the gap symmetry in (1) up to
the third-order term, we insert the pair potential (the second result of 
equation (5) )
and the corresponding gap function into
the BCS gap equation, $\Delta (k) = \sum_k^\prime (E_{k^\prime})^{-1}$
$V_{k,k^\prime} \Delta (k^\prime)\tanh(\beta E_{k^\prime} /2)$. Then, comparing
the $k$-dependence from both sides of the gap equation one gets the required
gap equation that produces the approximated spin fluctuation (ASF) gap symmetry
is obtained as,
\begin{eqnarray}
\Delta(0) = \sum_{k^\prime} (V/3) F_{k^\prime}^2 \frac{\Delta(0)}
{2 E_{k^\prime}}
\tanh (\frac{\beta E_{k^\prime}}{2}).
\end{eqnarray}

This is exactly what the gap equation one would find using the pair potential
$V(k,k^\prime) = V F_k F_{k^\prime}$ which certainly produces the required
gap symmetry $\Delta(k) = \Delta(0)f_k(1+g_{k}+g_{k}^2)$ (the pair vertex
in (7) is only renormalized to $V/3$). The symbols in equation (7)
have their usual meanings with the superconducting quasiparticle energy is
given by,
$E_k = \sqrt{(\epsilon_k -\mu)^2 + \Delta^2 (k)}$, where $\mu$ is the chemical
potential which controls the band filling with the help of a number
conserving equation. The temperature dependence of the chemical potential
is taken care in the self-consistent numerical solutions of the gap equation.
We use the band dispersion $\epsilon_k$ obtained from the angle resolved
photoemission experiment carried out 
by Norman {\it et al,} \cite{13} for the $Bi-$ based
cuprates. The dispersion gives experimentally measured hopping amplitudes
up to fifth neighbours which is plausible in the present calculation since
further neighbours attractive pairing interaction is considered.

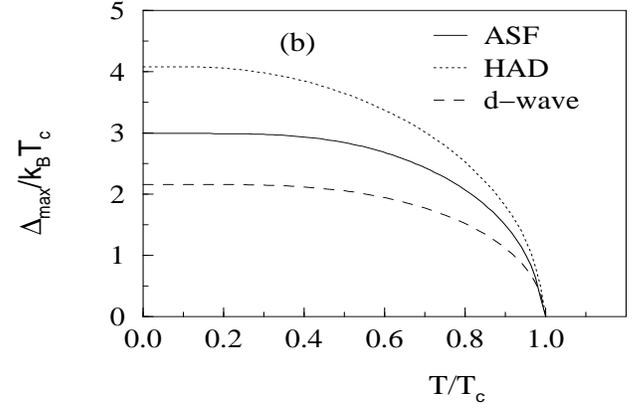
\begin{figure}
\epsfxsize=4.5truein
\epsfysize=4.5truein
\begin{center}
{\epsffile{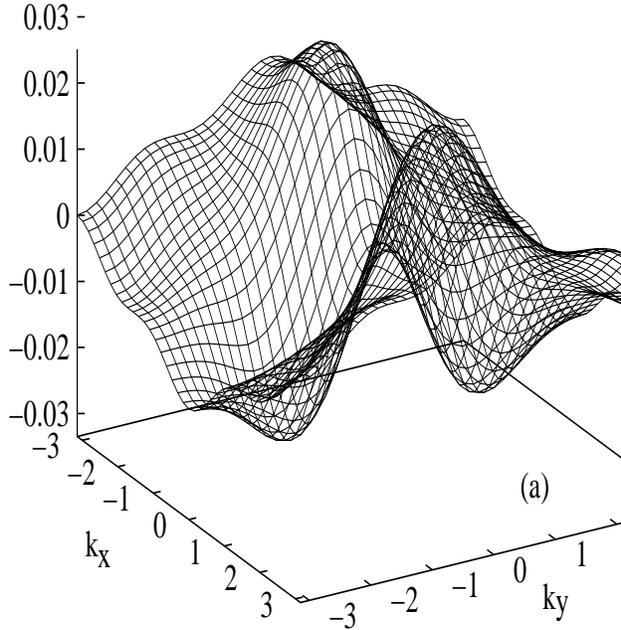}}
\epsfxsize=3.0truein
\epsfysize=2.5truein
{\epsffile{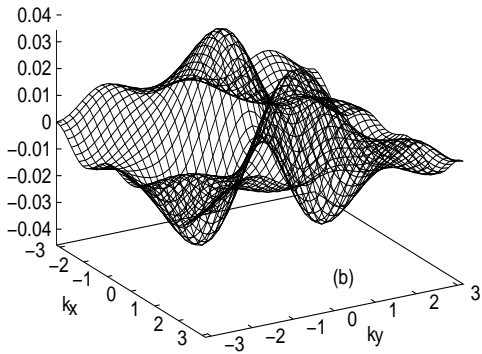}}
\caption{Momentum dependence of the superconducting energy gap function which
has symmetry of that of the approximate spin fluctuation (ASF)
theory (up to $3^{rd}$ order
in (1)). The required
attractive potential is derived within the approximation (6) {\bf (a)}. Note the
strong resemblence of this gap structure with that obtained by Lenck and
Carbotte (cf. figure 1 of ref. \cite{lenck}). This higher anisotropic $d$-wave
gap function yields a $2 \Delta(k)_{max}/k_BT_c= 6$ where the
$\Delta (k)_{max}$ is not at $(k_x,k_y) = (0,\pm \pi)$ but around
 $(0, \pm 1.57)$.
{\bf (b)} Momentum dependence of the superconducting energy gap function 
when the attractive potential is derived exactly without the approximation (6)
leading to higher anisotropic $d$-wave (HAD) symmetry.
Contrast, the deviation in the gap structure due to the approximation (6)
in {\bf (a)}. These gap structures {\bf (a, b)} are indeed consistent
 with that obtained by Lenck and
Carbotte (cf. \cite{lenck}). 
The exact HAD symmetry gap structure {\bf (b)} although have similar structure 
to ASF symmetry, the gap becomes more sharply peaked in different 
$k$ directions. Unlike the simplest $d$-wave its maximum occur around
 $(0,\pm 1.41)$ which yields a $2 \Delta(k)_{max}/k_BT_c= 8.15$,
the value very much consistent with the original spin fluctuation model 
\cite{9}. 
The parameter $V$ is
such adjusted that both the gap functions
have the same bulk $T_c = 84$ K at the band-filling
$\rho= 0.8$.
Undoubtedly, these gap structures are
 quite different from the {\em lowest order}
usual $d$-wave symmetry usually considered in the literature.}
\end{center}
\end{figure}

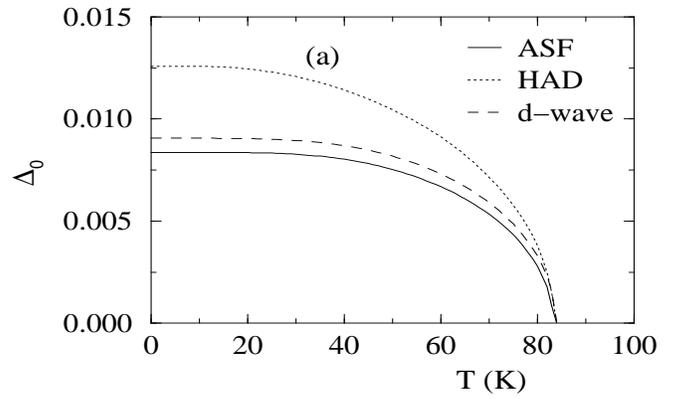
\begin{figure}
\epsfxsize=3.5truein
\epsfysize=4.20truein
{\epsffile{fig2}}
\caption{{\bf (a)}
Variation of the gap amplitude ($\Delta (0)$ in eV, 
see equations (1), (7), (8)) as
a function of temperature in Kelvin.
Note, the amplitude $\Delta (0)$ for the usual $d$ wave and the ASF (under 
approximation (6)) is almost same whereas the HAD 
(without the approximation (6)) is quite different. The gap opens up very
fast below $T_c$ in the HAD case. 
{\bf (b)}
The BCS characteristic ratio $\Delta (k)_{max}/k_BT_c$ as a
function of $T/T_c$ in different models. The solid
curve represents result for the derived approximated
spin fluctuation (ASF) symmetry within
the approximation (6); the maximum of the 
gap opens up at a faster rate than the $d$-wave below $T_c$. The dotted
line corresponds to the HAD symmetry 
when no approximation (6) is included; this
maximum gap has the fastest growth with lowering in temperature
below $T_c$.
The dashed curve corresponds to a usual {\em lowest order} $d$-wave.
Therefore, the special momentum anisotropic form in the spin fluctuation theory
[Monthoux, Balatsky, Pines, Phys. Rev. Lett. {\bf 67}, 3448] is one of 
the crucial ingredients for so high BCS characteristic ratio. 
(All the curves in the figure
correspond to $T_c = 84$ K at the band filling $\rho = 0.8$).
}

\end{figure}
 Following the same principle as in deriving equation (7), we get the 
gap equation
for the higher anisotropic $d$ wave (HAD) symmetry $\Delta (k) = 
\Delta (0) [f_1(k)+f_3(k)+f_{4a}(k)+f_{4b}(k)+f_6(k)]$ using the full
potential {\it i.e,} the first result of the equation (5) as,
\begin{equation}
\Delta(0) = \sum_{k^\prime} (V/5) \tilde F_{k^\prime}^2 \frac{\Delta(0)}
{2 E_{k^\prime}}
\tanh (\frac{\beta E_{k^\prime}}{2})
\end{equation}
where $\tilde F_{k} = f_1(k)+f_3(k)+f_{4a}(k)+f_{4b}(k)+f_6(k)$ with
$f_1(k) = (1/\sqrt{2})f_k$, $f_3(k)=f_kg_k$, $f_{4a}(k)=(1/\sqrt{2})f_k
(1+s_{k_{xy}})$, $f_{4b}(k)=(1/\sqrt{2})f_kd_{k_{xy}})$,
$f_6(k) =f_k(g_{k}^2-s_{k_{xy}}/2-3/4)$.
Now, we present our numerical results in figures 1 $\&$ 2, for a fixed
cut-off frequency $\Omega_c = 500$ K. The bulk $T_c$ in all the figures is
fixed at $T = 84$ K for the band filling $\rho = 0.8$,
which required a change in $V$ in the  superconducting gap equations (7), (8)
and the same for the usual $d$-wave. 
In figures 1 (a,b) we present the $k$-anisotropy of the gap function in the
first Brillouin
zone for $k_x$, $k_y$, which clearly produces $d$-wave like
solution {\it i.e,} nodal lines along $k_x = k_y$ directions
 as well as change in
the sign of the gap function. However, the overall anisotropy is very
different from the simplest {\em lowest order} $d$-wave ($\Delta (k) =$
$\Delta (0) (\cos k_x a - \cos k_y a)$) form \--- 
but rather a {\em higher anisotropic
$d$-wave}. Figure 1(a) presents the momentum anisotropy for the ASF model
using gap equation (7) whereas the Figure 1(b) represents the HAD symmetry
 using
the gap equation (8). The HAD symmetry gap has sharper k-anisotropy than the
ASF model although there is overall similarity between the
two, namely, positions of maximum gap are very close. Since the calculation of
Lenck and Carbotte \cite{lenck} does not assume any form of the
superconducting gap function and also does not include retardation effects,
as in the original spin fluctuation model \cite{9}, we can therefore 
certainly rely on
comparing
our results with those of Lenck and Carbotte. 
A close comparision of our results
with those of \cite{lenck} will conclusively demonstrate that 
the pair potential
derived with distant neighbours attraction in the $d$-wave channel 
in the present model {\em does} produce the 
gap symmetry of the spin fluctuation theory \cite{9}.
The present calculation thus may indicate that 
 the phenomenological spin fluctuation theory includes longer range
interaction which might be derivable from a generalised interaction (2).

In figure 1, the gap function shows more than one maximum (minimum) at the
edges of the Brillouin zone, very similar to that obtained in
reference \cite{lenck}. The maximum (minimum) is also displaced from the usual
position ($(0,\pm \pi), (\pm \pi, 0)$) in simplest $d$-wave (cf. figure 1 and
its caption). Remarkably, this
gap symmetry also produces high value of $2 \Delta (k)_{max}/k_B T_c = 6$ same
as that obtained in \cite{lenck} (cf. figure 2(b)). In Fig.1(b)
 where we present the same as
in Fig.1(a) but without the approximation (6) the BCS characteristic ratio
$2 \Delta (k)_{max}/k_B T_c = 8.15$. These are values close to typical of
what is known for high-$T_c$ systems \cite{9}. 
Note, however, there may be some differences between the two (work 
\cite{lenck} and this one)
only in {\em details}, because the band dispersion used in the two calculations
are different which does effect pairing symmetry.

Having discussed in details the difference in the $k$-anisotropy of the 
(higher anisotropic $d$-wave) HAD 
symmetry and that of the usual $d$-wave, and shown 
that such symmetries {\em do}
reproduce the gap structure of the spin fluctuation theory, we now show
explicitly in figures 2 (a,b) that due to strong anisotropy such gaps have
different thermal behavior in comparison to the usual $d$-wave which is 
principle cause for high value of $2 \Delta (k)_{max}/k_B T_c$.
In Fig. 2(a) the amplitudes of higher anisotropic $d$-wave symmetries 
(with and
without approximation (6)) and that of the simple $d$-wave are displayed 
as a function of temperature (T) \---- all of them have 
same $T_c = 84$ K at a band-filling $\rho = 0.8$. In figure 2(b) we pick up 
the temperature dependencies
 of the maximum gap of the three $d$-wave symmetries same as shown in
figure 2(a) and display the $\Delta (k)_{max}/k_B T_c$ ratio 
as a function of the
reduced temperature $T/T_c$. The gap opens up below $T_c$ at the fastest 
rate for the HAD symmetry and at the slowest
rate for the usual $d$-wave as the temperature is lowered.

 Finally, to summarize, we have derived a pair potential from further
neighbours attraction in the tight binding scenario which produces gap symmetry
of the phenomenlogical spin fluctuation theory.
This study may  particularly be
justified from the fact that in models of spin fluctuation
mediated $d$-wave superconductivity an increase in the antiferromagnetic
correlation length occurs with underdoping.
Such effect has also been realized very recently from angle resolved
photoemmission (ARPES) experiment by a well known group \cite{mesot}. One
of their principle observations is that as the doping decreases the maximum
gap increases, but the slope of the gap near the nodes decreases. This
feature, although consistent with $d$ wave symmetry, cannot be fitted using
 the
simplest $d$-wave form of the gap but requires a more generalised $d$-wave
of the form,
$B(\cos k_x-\cos k_y)+(1-B)(\cos 2k_x -\cos 2k_y)$ where $B$ is a fitting 
parameter. (Needless to say, in the present work consideration of
only first two terms of the first result in equation (5) {\it i.e}, 
$1^{st}$ and $3^{rd}$ neighbour interactions exactly reproduces this symmetry).
 This lead them to suggest importance of longer range
interaction in the theory of $d$-wave superconductivity as
one approaches the insulator.
It is worth pointing out that such ARPES experiments in the underdoped
regime measure the
pseudo-gap rather than the truly superconducting gap.
In the present calculation, a close look to the dispersion 
used in equations (7,8) will 
indicate that the Fermi surface (FS) is open in certain direction {\it i.e,}  
the FS is gapped due to pseudo-gap formation.
 However, if the 
pseudo-gap could be ascribed to fluctuation effects of the order
parameter, then its value could be estimated by the mean-field BCS equation,
while the truly superconducting transition can be estimated only by 
calculations that include flutuation effects. 
 We thus created an example that
there exist in nature pair potential (as we derive in real space)
 analogous to the
spin fluctuation theory which is one of the leading potential theories
in high temperature superconductors, despite the fact that
 the principle philosophy of the spin
fluctuation is different. 
 We thus emphasized, the importance of
inclusion of further
neighbour attraction in the usual $d$-wave theories as is also realized
in most recent experiment \cite{mesot}. 
The $Cu-O$ systems
being in a complicated circuit, the effect of columb repulsion may not be
adequately treated with only on-site repulsion and therefore,
effective attractive
potential may be achieved only after considering more distant
 neighbours terms.
With this calculation we also emphasized the role of gap anisotropy in
BCS gap ratio which may be further improved adopting a strong coupling
approach \cite{14}. 
The higher anisotropic $d$-wave symmetry as obtained in this
work will certainly be consistent with experimental studies in cuprates because
of its similarity with $d$-wave but will have better advantage of avoiding
coulomb repulsion. We believe, this work along with \cite{9,nmr,lenck,mesot}
will provide new insight to the usual $d$-wave theories of superconductivity.

The author
 thanks Sreekantha Sil for stimulating communications and useful comments
 via electronic mail on the subject.
This work was partly
 supported by the Brazilian Funding Agency FAPERJ, project no.
E-26/150.925/96-BOLSA.

\end{document}